\documentclass[12pt]{article}	
\pdfoutput=1 
\usepackage{cite}
\usepackage{color} 
\usepackage{amsmath}
\usepackage{amsfonts}
\usepackage{amssymb}
\usepackage{array,booktabs}
\usepackage{float}
\usepackage{caption}
\usepackage{subcaption}
\usepackage{graphicx}
\usepackage{slashed}            
\usepackage{bbm}                
\usepackage{xspace}				
\usepackage{collref}				
\usepackage{framed}				
\usepackage{placeins}
\usepackage{jheppub}							
\usepackage{mathrsfs}
\usepackage{hyperref}
\usepackage{cancel}
\usepackage[normalem]{ulem}

\newcommand{\beq}{\begin{equation}}
\newcommand{\eeq}{\end{equation}}
\newcommand{\beqa}{\begin{equation}\begin{aligned}}
\newcommand{\eeqa}{\end{aligned}\end{equation}}
\newcommand{\lsim}{\lesssim}
\newcommand{\gsim}{\gtrsim}

\newcommand{\Eq}[1]{(\ref{eq:#1})}
\newcommand{\Fig}[1]{Fig.~\ref{fig:#1}}
\newcommand{\Ref}[1]{Ref.~\cite{#1}}
\newcommand{\Sec}[1]{Sec.~\ref{sec:#1}\@}

\newcommand{\GeV}{\>\mathrm{GeV}}
\newcommand{\Mpc}{\>\mathrm{Mpc}}
\newcommand{\Neff}{\Delta N_\text{eff}}
\newcommand{\dd}{\mathrm{d}}
\newcommand{\cO}{\mathcal{O}}
\newcommand{\DM}{{\text{\tiny DM}}}
\newcommand{\DR}{{\text{\tiny DR}}}
\newcommand{\B}{{\text{\tiny B}}}

\begin{document}
\title{Partially Acoustic Dark Matter, Interacting Dark Radiation, and Large Scale Structure}

\author{Zackaria Chacko$^{1}$, Yanou Cui$^{1,2,3}$, Sungwoo Hong$^{1}$, Takemichi Okui$^{4}$, and Yuhsin Tsai$^{1}$}
\affiliation{$^{1}$Maryland Center for Fundamental Physics, Department of Physics, University of Maryland, College Park, MD 20742, USA\\
$^{2}$Department of Physics and Astronomy, University of California-Riverside, Riverside, CA 92521, USA\\
$^{3}$Perimeter Institute, 31 Caroline Street, North Waterloo, Ontario N2L 2Y5, Canada\\
$^{4}$Department of Physics, Florida State University, Tallahassee, FL 32306, USA
}


\abstract{
The standard paradigm of collisionless cold dark matter is in tension with measurements on large scales. In particular, the best fit values of the Hubble rate $H_0$ and the matter density perturbation $\sigma_8$ inferred from the cosmic microwave background seem inconsistent with the results from direct measurements. We show that both problems can be solved in a framework in which dark matter consists of two distinct components, a dominant component and a subdominant component. The primary component is cold and collisionless. The secondary component is also cold, but interacts strongly with dark radiation, which itself forms a tightly coupled fluid. The growth of density perturbations in the subdominant component is inhibited by dark acoustic oscillations due to its coupling to the dark radiation, solving the $\sigma_8$ problem, while the presence of tightly coupled dark radiation ameliorates the $H_0$ problem. The subdominant component of dark matter and dark radiation continue to remain in thermal equilibrium until late times, inhibiting the formation of a dark disk. We present an example of a simple model that naturally realizes this scenario in which both constituents of dark matter are thermal WIMPs. Our scenario can be tested by future stage-IV experiments designed to probe the CMB and large scale structure.
}
\maketitle

\section{Introduction}
\label{sec:intro}
For nearly two decades, the $\Lambda$CDM paradigm in which dark matter 
(DM) is composed of cold, collisionless particles has provided an 
excellent fit to cosmological data. Although on galactic scales or 
smaller there have been long-standing issues such as the 
core-vs-cusp~\cite{Moore:1994yx, Flores:1994gz} and 
too-big-to-fail~\cite{2011MNRAS.415L..40B} problems that are difficult 
to explain within this framework, it has been very successful on larger 
scales. However, in recent years, as the data has become more precise, 
the $\Lambda$CDM framework has also come into tension with measurements 
on large scales. In particular, the value of today's Hubble rate $H_0$ 
obtained from a fit to the cosmic microwave background (CMB) and baryon 
acoustic oscillation (BAO) data \cite{Ade:2015xua} is smaller than the results from local 
measurements~\cite{Riess:2011yx, 2013ApJ...766...70S, Riess:2016jrr,DiValentino:2016hlg, Bernal:2016gxb}, 
with a $\sim 3\sigma$ discrepancy. Similarly, the inferred value of 
$\sigma_8$ (the amplitude of matter density fluctuations at the 
scale of $8h^{-1}\Mpc$) is larger by 
$3$--$4\sigma$~\cite{Heymans:2013fya, Ade:2013lmv, MacCrann:2014wfa} 
than the values from direct measurements such as weak lensing 
survey~\cite{Fu:2014loa}, CMB lensing~\cite{Ade:2015zua}, and 
Sunyaev-Zeldovich cluster counts~\cite{2012ApJ...755...70R, 
Hasselfield:2013wf, Ade:2015fva}.

These large-scale anomalies are particularly intriguing because the 
theoretical understanding of dynamics at large scales is rather simple 
and robust, essentially requiring only the application of linear 
perturbation theory to density fluctuations. This is in contrast to the 
studies of small scale structure, which not only require an 
understanding of the nonlinear evolution of a many-body system but also 
crucially depend on the detailed dynamics of baryons, which is quite 
challenging to simulate. Therefore, while it is possible that these large scale 
discrepancies are due to systematic errors in the associated 
experiments~\cite{Joudaki:2016mvz,Kitching:2016hvn}, it is important to 
consider the possibilty that they are in fact robust problems that 
require a fundamental shift away from the $\Lambda$CDM framework.

Several proposals have been put forward to explain one or the other of 
these anomalies by going beyond the $\Lambda$CDM paradigm. For example, 
DM that decays at late times, well after the CMB epoch, can reduce the 
size of $\sigma_8$~\cite{Enqvist:2015ara,Poulin:2016nat}. Alternatively, neutrinos with masses near the top of the allowed range or sterile neutrinos with an eV mass can fit the 
CMB and BAO data with a smaller 
$\sigma_8$~\cite{Battye:2014qga}, but have 
the effect of making the $H_0$ problem worse~\cite{Ade:2015xua}. If there is dark radiation 
(DR) that behaves like a tightly coupled fluid (as opposed to free 
streaming like neutrinos), this can ameliorate the tension in $H_0$ 
measurements~\cite{Baumann:2015rya}. If DM further scatters with such 
DR, it may be possible to solve both the $H_0$ and $\sigma_8$ 
problems~\cite{Buen-Abad:2015ova, Lesgourgues:2015wza, Ko:2016uft, Ko:2016fcd}. 
These proposals, however, require a rather tiny DM-DR interaction 
constrained in a very narrow range to solve both problems.

In this article we propose a new simple framework, ``Partially Acoustic 
Dark Matter'' (PAcDM), that can robustly solve both problems. We assume 
that DM consists of two components, $\chi_1$ and $\chi_2$, and that 
there is also DR that behaves as a tightly coupled relativistic 
fluid.\footnote{Earlier work on multi-component DM may be found in, for 
example,~\cite{Goldberg:1986nk, Khlopov:1989fj, Berezhiani:1995am, 
Kaplan:2009de, Kaplan:2011yj, CyrRacine:2012fz, Dienes:2011ja}.} The 
primary component $\chi_1$ is cold and collisionless, and dominates the 
DM mass density. The subdominant component, $\chi_2$, is also cold, but 
is tightly coupled to the DR\@. The interactions within the DR, and 
between the DR and $\chi_2$, are both strong enough that the tight 
coupling treatment is valid not only during radiation domination before 
the CMB epoch but also well into the era of structure formation. Then, 
since our DR is a tightly coupled relativistic fluid as considered in 
\Ref{Baumann:2015rya}, we can solve the $H_0$ problem by choosing the 
amount of DR appropriately. We will then demonstrate that the persistent 
$\chi_2$-DR interaction inhibits the growth of density perturbations in 
$\chi_2$, which in turn reduces the 
growth of density fluctuations in the dominant DM component, $\chi_1$, 
provided that the modes in question enter the horizon before 
matter-radiation equality. The modes at the $8h^{-1}\Mpc$ scale do 
indeed come inside the horizon before equality, so we can also solve the 
$\sigma_8$ problem just by choosing an appropriate amount of $\chi_2$ to 
match the observed discrepancy in $\sigma_8$. 

This class of theories fits very naturally into a ``hidden WIMP" 
scenario~\cite{Finkbeiner:2007kk, Pospelov:2007mp, Feng:2008ya, 
Feng:2008mu}, in which 
the relic abundance of both DM components $\chi_1$ and $\chi_2$ is set 
by annihilation into a hidden sector, rather than into the SM. Massless 
states in this hidden sector could then constitute interacting 
DR~\cite{Chacko:2015noa}, ameliorating the $H_0$ problem, while 
scattering of $\chi_2$ off DR solves the $\sigma_8$ problem. In the next 
section, we construct an explicit model along these lines. However, we 
stress that the qualitative features of our scenario---an enhancement in 
$H_0$ and a reduction in $\sigma_8$---are robust and only require the 
coupling constants in the DR-$\chi_2$ sector to be sufficiently large 
that DR is a tightly coupled fluid and $\chi_2$ remains in equilibrium 
with DR\@. The mechanism is therefore quite general, and is not 
restricted to a specific DM framework.

The PAcDM paradigm shares some features of the ``double-disk dark 
matter'' scenario explored in~\cite{Fan:2013yva, Fan:2013tia}, but there 
are several crucial differences. In particular, since our ``dark 
electrons'' are massless, they never form bound states and, 
consequently, the $\chi_2$-DR system never undergoes recombination. 
Therefore, after matter-radiation equality, it continues to undergo 
\emph{dark acoustic oscillations}~\cite{Cyr-Racine:2013fsa} without 
being disrupted by ``dark recombination.'' This has the effect of 
holding back the growth of density perturbations in $\chi_2$ until the 
energy density in DR falls too much for the oscillations to be 
maintained.  Furthermore, by assumption, the $\chi_2$-DR system remains 
tightly coupled throughout the evolution of the universe, even into the 
era of structure formation. At later times it continues to remain a 
\emph{thermal system}, and hence does not virialize. Therefore, we expect that 
it does not collapse into a disk but instead forms a spherical halo 
around the galactic center.

The rest of this paper is organized as follows. In \Sec{model}, we 
describe a concrete, complete model that realizes our solution to the 
$H_0$ and $\sigma_8$ problems within the hidden WIMP framework. In 
\Sec{evolution}, we develop an analytical understanding of how the 
$\sigma_8$ problem is solved in the PAcDM scenario by studying the linear 
evolution equations for the density and metric perturbations. This is 
done without assuming any particular particle physics model of the DM 
and DR\@. In \Sec{numerical}, we present the results of our more 
detailed numerical simulations of the matter power spectrum to 
determine the precise fraction of $\chi_2$ to match the observed 
discrepancy in $\sigma_8$.  We then show that the corrections to the CMB 
spectrum and the CMB lensing measurement are small and within the 
present uncertainties. We conclude in \Sec{conc}.

\section{A Benchmark Model} 
\label{sec:model} 
In this section, we construct a benchmark model based on the hidden 
WIMP paradigm that realizes these ideas. The hidden WIMP framework, in 
which DM annihilates into hidden sector states rather than into SM 
particles, has become increasingly attractive as the constraints from 
direct, indirect, and collider searches continue to get stronger. This 
class of theories retains the attractive feature of the WIMP paradigm 
that the DM mass and annihlation cross section are of order the weak 
scale, a scale that is motivated by a variety of solutions to the 
hierarchy problem. In the model we consider, both $\chi_1$ and $\chi_2$ 
are hidden WIMPs, and their abundances are set by annihilation within the hidden sector.

In a hidden WIMP model the couplings between DM and the SM states may be 
very small. Nevertheless, as emphasized in~\Ref{Chacko:2015noa}, if the 
temperature of the hidden sector is comparable to that of the SM, this 
class of theories is potentially accessible to experiment. The nature of 
the signals depends on the masses of the lightest states in the hidden 
sector. If all the particles in the hidden sector have masses above an 
eV, these states must decay or annihilate into SM particles before the 
CMB epoch. This is so as to avoid the over-closure bounds if their 
masses are above a keV, or the limits on a warm subcomponent of DM if 
their masses lie between an eV and a keV\@. This then implies the 
existence of couplings between the hidden sector and the SM that can 
potentially be tested in experiments, as in the exciting 
DM~\cite{Finkbeiner:2007kk,Cholis:2008vb}, secluded 
DM~\cite{Pospelov:2007mp}, and boosted DM~\cite{Agashe:2014yua, 
Berger:2014sqa} scenarios. If instead the lightest states in the hidden 
sector have masses below an eV, they are likely to constitute a 
significant component of the energy density of the universe at the time 
of matter-radiation equality, potentially leading to observable signals 
in the CMB. The simplest possibility is that these states, if present, 
are massless and constitute DR at present times. Provided the hidden 
sector was in thermal equilibrium with the SM at temperatures at or 
above the weak scale, there is a lower bound on the contribution of DR 
to the energy density during the CMB epoch, $\Delta N_{\rm eff} \gsim 
0.027$~\cite{Chacko:2015noa}, as expressed in units of the energy density 
in a single SM neutrino species. This is potentially large enough to be detected in 
CMB Stage-IV experiments~\cite{Abazajian:2013oma, 
Errard:2015cxa, Wu:2014hta, Dodelson:2016wal}. We see that the existence of observable DR is 
a natural feature in a large class of hidden WIMP models.

DR can take two distinct forms; it may free stream like neutrinos, or 
scatter with a short mean free path like the particles in a tightly 
coupled fluid. These two cases can be experimentally distinguished, 
because the details of the CMB spectrum depend not just on the total 
energy density in radiation, but on the fraction that is free 
streaming~\cite{peebles1973role, Hu:1995en, Bashinsky:2003tk}. This ratio impacts both 
the amplitudes of the modes, and the locations of the peaks in the CMB 
spectrum. The effects of scattering DR and free streaming DR on the CMB 
are therefore different. In the case when DR scatters, a fit to the 
Planck CMB data admits values of $\Delta N_{\rm eff}$ large enough to 
address the $H_0$ problem~\cite{Baumann:2015rya}. As we shall see, 
scattering of $\chi_2$ off the DR can address the $\sigma_8$ problem. In 
this way the hidden WIMP framework is well suited to address both the 
large scale puzzles.
 
A concrete, renormalizable and radiatively stable field theory model of 
a hidden WIMP that annihilates into interacting DR was presented and 
analyzed in~\Ref{Chacko:2015noa}. In this theory, the DR consists of one 
or more flavors of massless Dirac fermions $\hat{\psi}$ that carry 
charge under a new U(1) gauge symmetry. The massless gauge boson associated 
with this symmetry is denoted by $\hat{A}$. Provided that the U(1) gauge 
coupling is not too small, $\hat{A}$ and $\hat{\psi}$ behave as tightly 
coupled DR at late times, and can solve the $H_0$ problem.
 
In the model of~\Ref{Chacko:2015noa}, the DM was composed of a single 
component, a complex scalar $\chi$, whose relic abundance was governed 
by the annihilation through a massive vector boson $\hat{Z}$ into the 
dark fermions $\hat{\psi}$. It is straightforward to extend this model 
by introducing a second component of DM, so that we now have two complex 
scalars, $\chi_1$ and $\chi_2$. The relic abundances of both $\chi_1$ 
and $\chi_2$ are now set by annihilation to the $\hat{\psi}$ through the 
massive vector boson $\hat{Z}$. A standard thermal freeze-out 
calculation then tells us that, in such a scenario, the ratio of their 
relic abundances is given by
\beq 
\frac{\Omega_2}{\Omega_1} \sim \!\left( \frac{m_2 q_1}{m_1 q_2} \right)^{\!\! 2} \,,
\label{eq:Omega2/Omega1} 
\eeq 
 where $m_{1,2}$ and $q_{1,2}$ are the mass and $\hat{Z}$ charge of 
$\chi_{1,2}$, respectively. We will see that a value of this ratio of 
order a few \% is required to match the observed $\sim 10\%$ discrepancy in $\sigma_8$. As is clear 
from the above expression, DM masses and $\hat{Z}$ charges that differ 
by just a factor of a few can easily result in such a value.
The SM and dark sectors can be arranged to be in thermal equilibrium at 
early times through small Higgs portal couplings $|\chi_{1,2}|^2 
|H|^2$, ensuring that their temperatures are not very different at later times.
This model shows that the thermal WIMP framework can provide a simple 
realization of our solution to the $H_0$ and $\sigma_8$ problems, 
without the need to introduce new scales or small parameters.

We now estimate the range of parameters for which this model gives rise 
to the desired dynamics. First, to solve the $\sigma_8$ problem, we want 
$\chi_2$ to be in thermal equilibrium with DR until late times, 
including the era of structure formation. This requires the momentum 
transfer rate between the two systems to be much larger than the Hubble 
rate $H$ until late times. To be quantitative, we define the momentum 
transfer rate $\Gamma$ from DR to $\chi_2$ per degree of freedom of 
$\chi_2$ as
\beq
\Gamma \equiv 
\frac{1}{\langle p_2^2 \rangle} \frac{\dd \langle \delta p_2^2 \rangle}{\dd t}
\,,\label{eq:Gamma_def} 
\eeq
where $\delta \vec{p}_2$ is the change of momentum of $\chi_2$ due to a 
single scattering event with a degree of freedom in DR, and the 
$\langle~\rangle$ indicates thermal average with DR being treated as the 
bath. Here $t$ is simply the Minkowski time, because $\Gamma$ is 
a microscopic quantity independent of the cosmological expansion, and can 
be determined in the flat spacetime limit. An elementary calculation shows 
$\Gamma \propto \hat{\alpha}^2 \ln(\hat{\alpha}^{-1}) \, 
\hat{T}^2/m_{\chi_2}$ \cite{Buen-Abad:2015ova},
%
%
where $\hat{\alpha}$ is 
the dark fine structure constant and $\hat{T}$ is the temperature of the 
dark sector. The proportionality factor is $\cO(1)$ and not important, 
provided that all the dark charges involved in the scattering are 
$\cO(1)$. On the other hand, the Hubble rate $H$ scales as $T^2$ during 
the radiation dominated era, but as $T^{3/2}$ during the matter 
dominated era. Since we want $\Gamma \gg H$ at all times, it follows that the 
strongest bound comes from the latest times. Demanding $\Gamma \gg H_0$ 
then leads to
\beq
\hat{\alpha}^2 \ln(\hat{\alpha}^{-1}) \, \frac{10\GeV}{m_{\chi_2}} 
\gg 10^{-16} \, \biggl( \frac{T_0}{\hat{T}} \biggr)^{\!\! 2}
\,,\label{eq:Gamma>>H0}
\eeq
where $T_0$ is the photon temperature today. It follows that we can 
easily maintain the thermal equilibrium between $\chi_2$ and DR at all 
times even with fairly small values of the dark gauge coupling. We also 
need to maintain the equilibrium within DR itself, but this clearly 
leads to a much weaker condition than the above because the relevant 
momentum transfer rate simply scales as $\hat{\alpha}^2 \hat{T}$, 
without the additional $\hat{T}/m_{\chi_2}$ suppression. Finally, note 
that Eq.~\Eq{Omega2/Omega1} is only valid when the relic abundances of 
$\chi_{1,2}$ are set by the $\hat{Z}$ gauge coupling, and not by 
$\hat{\alpha}$. This leads to the condition $\hat{\alpha} \ll 
\hat{\alpha}_{\hat{Z}} \,(m_{\chi_{1,2}} / m_{\hat{Z}})^2$. (If 
$\hat{\alpha}$ is larger than this, the annihilation through $\hat{A}$ 
dominates and Eq.~\Eq{Omega2/Omega1} would have to be modified 
accordingly.) With a typical size of $\hat{\alpha}_{\hat{Z}}$ for 
WIMPs, we see that this leaves a wide range of parameters for which the 
$\sigma_8$ problems is solved.

Second, in order to solve the $H_0$ problem, we require 
$\Neff^\mathrm{scatt} \gtrsim 0.4$~\cite{Riess:2016jrr,Bernal:2016gxb}, where 
$\Neff^\mathrm{scatt}$ is the energy density in the scattering component 
of DR, measured in units of the energy density in a single SM neutrino 
species. The size of $\Neff^\mathrm{scatt}$ in the concrete model being 
discussed in this section can be read off from \Ref{Chacko:2015noa}. For 
example, if the SM and dark sectors kinetically decouple from each other 
between the QCD phase transition temperature and the $c$-quark mass 
(which can easily be arranged to be the case), we get 
$\Neff^\text{scatt} \simeq 0.25$ (or $0.42$) for one (or two) flavor(s) 
of $\hat{\psi}$ and $\hat{T} \simeq 0.39 \,T_0$. These values of 
$\Neff^\text{scatt}$ are allowed by the current CMB 
constraints~\Ref{Baumann:2015rya, Lesgourgues:2015wza, Nscatt}, allowing 
a straightforward solution of the $H_0$ problem.

\section{General evolution of DM density perturbations}
\label{sec:evolution}
 In this section, we demonstrate how the $\sigma_8$ problem is robustly 
solved in our framework by performing a simple analysis of the 
perturbed Einstein equations. More detailed, numerical studies will be 
presented in the next section. Our analysis here is very general, 
and essentially depends on only the following assumptions: 
 \begin{itemize}
 \item{There is a significant contribution to the energy density of the 
universe from DR around the time of
matter-radiation equality. This DR is tightly self-coupled not only during the 
CMB epoch but also at late times when large scale structure forms.}
 \item{DM is cold and comes in two types, a dominant component $\chi_1$ 
that is collisionless, and a subdominant component $\chi_2$ that 
interacts strongly with the DR, again until late times.}
 \end{itemize}
 To study how the growth of the perturbations of $\chi_1$ is reduced by the 
scattering of $\chi_2$ with DR, we employ the general formalism of Ma 
and Bertschinger~\cite{Ma:1995ey} for scalar perturbations. (The study 
of tensor perturbations is beyond the scope of the present work.) We 
work in the conformal Newtonian gauge,
\beq
\dd s^2 
= a^2(\tau) \, 
  \bigl[ -(1 + 2 \psi) \dd\tau^2 + (1 - 2\phi) \delta_{ij} \dd x^i \dd x^j \bigr] \,,
\eeq
where the fields $\psi$ and $\phi$ describe scalar perturbations on the 
metric. They are determined by four scalar quantities associated with 
the perturbed energy-momentum tensor $\delta T_\mu^\nu$, namely, $\delta 
\equiv \delta \rho / \bar{\rho} = -\delta T_0^0 / \bar{\rho}$, $\delta P 
= \delta T_i^i / 3$, $\theta \equiv -\partial_i \delta T_0^i / 
(\bar{\rho} + \bar{P})$, and $\sigma \equiv -\hat{\partial}_i 
\hat{\partial}^j (\delta T^i_j - \delta P \, \delta^i_j) / (\bar{\rho} + 
\bar{P})$, where $\bar{\rho}$ and $\bar{P}$ are the unperturbed total 
energy density and pressure, and $\hat{\partial}^i \equiv 
\hat{\partial}_i \equiv \partial_i / \sqrt{\partial_j \partial_j}$. For 
each particle species $s$, we define $\delta_s \equiv \delta\rho_s / 
\bar{\rho}_s$, $\theta_s \equiv -\partial_i \delta T_{s0}^i / 
(\bar{\rho}_s + \bar{P}_s)$, etc. We can also re-express $\theta_s$ as 
the divergence of comoving 3-velocity, $\theta_s = \partial_i v^i_s$, 
where $v^i \equiv \dd x^i / \dd \tau$. $\sigma_s$ is the shear stress of 
the fluid component $s$ and vanishes if $s$ is a perfect fluid. For each 
$s$ we assume an equation of state of the form $P_s = w_s \rho_s$ with 
constant $w_s$, so the pressures and energy densities are not 
independent quantities. The total $\delta$ and $\theta$ are given in 
terms of the individual $\delta_s$ and $\theta_s$ as $\delta = \sum_s 
\bar{\rho}_s \delta_s / \bar{\rho}$ and $\theta = \sum_s (\bar{\rho}_s + 
\bar{P}_s) \theta_s / (\bar{\rho} + \bar{P})$. $\sigma$ is related to 
the individual $\sigma_s$ by a sum rule analogous to that for $\theta$. 
If the total $\sigma$ is zero, the two metric perturbations become 
equal, $\psi = \phi$. An important parameter in our framework is the 
ratio $r$ defined as
\beq
r \equiv \frac{\bar{\rho}_2}{\bar{\rho}_\DM}
\label{eq:r_def}
\eeq
with $\bar{\rho}_\DM \equiv  \bar{\rho}_1 + \bar{\rho}_2$.

To linear order in the perturbations, the evolution of the dominant, 
collisionless component of DM, $\chi_1$, is described by
\beq
\dot{\delta}_1 = -\theta_1 + 3\dot{\phi}
\,,\quad
\dot{\theta}_1 = -\frac{\dot{a}}{a} \theta_1 + k^2 \psi
\,,\label{eq:evolution:DM1}
\eeq
where the dots indicate derivatives with respect to $\tau$, and we have 
switched to Fourier space via the transformation 
$\displaystyle{\mathrm{e}^{\mathrm{i} \vec{k} \cdot \vec{x}}}$. These 
equations are just the expressions of local conservation of energy and 
momentum of $\chi_1$ in an expanding FRW background. Being collisionless 
and cold, $\chi_1$ evolves exactly as standard cold DM (CDM) for \emph{given} 
$\phi$ and $\psi$. But we will see that in the presence of scattering 
between $\chi_2$ and DR, $\phi$ and $\psi$ are modified away from their 
values in a standard, single-component CDM model. $\chi_2$ evolves as
\beq
\hspace{-0.6ex}
\dot{\delta}_2 = -\theta_2 + 3\dot{\phi}
\,,\quad
\dot{\theta}_2 = -\frac{\dot{a}}{a} \theta_2 + k^2\psi +
a\Gamma (\theta_\DR \!- \theta_2) 
\,,\label{eq:evolution:DM2}
\eeq
where $\Gamma$ is the rate of momentum transfer as defined 
in~\Eq{Gamma_def}, while DR evolves as
\beqa
\dot{\delta}_\DR &= -\frac43\theta_\DR + 4\dot{\phi}
\,,\\
\dot{\theta}_\DR &= k^2 \Bigl( \frac{\delta_\DR}{4} + \psi \Bigr) + \frac34 \frac{\bar{\rho}_2}{\bar{\rho}_\DR} a\Gamma (\theta_2 - \theta_\DR) 
\,.\label{eq:evolution:DR}
\eeqa
Equations~\Eq{evolution:DM2} and~\Eq{evolution:DR} also just express 
local energy-momentum conservation, but now conservation applies only to 
the total $\chi_2$-DR system. Accordingly, it may be observed that the 
$\Gamma$ terms cancel for the total $\theta \propto \bar{\rho}_2 
\theta_2 + \frac43 \bar{\rho}_\DR \theta_\DR$, where we have made use of 
the sum rule for $\theta$ described earlier. It is implicitly assumed 
in~\Eq{evolution:DM2} and~\Eq{evolution:DR} that all the constituents of 
DR (in our case, dark massless electrons and dark photons) are always 
sufficiently tightly coupled to each other that DR can be treated as a 
single fluid without a shear stress.

Now, one of our key assumptions is that the $\chi_2$-DR interaction 
rate, $\Gamma$, is also always large enough that the tight coupling 
approximation {can be applied to~\Eq{evolution:DM2} 
and~\Eq{evolution:DR} not only during the radiation dominated era but 
also throughout the matter dominated era. Then, expanding $\theta_2$, 
$\psi$, etc., in powers of $1 / \Gamma$ in~\Eq{evolution:DM2} 
and~\Eq{evolution:DR}, we get $\theta_2 = \theta_\DR$ to leading 
order. Then, cancelling the $\Gamma$ terms by taking an appropriate 
linear combination of the $\dot{\theta}_2$ and $\dot{\theta}_\DR$ 
equations and then setting $\theta_2 = \theta_\DR \equiv 
\tilde{\theta}$, we obtain
\beq
\dot{\tilde{\theta}}
= -\frac{\dot{a}}{a} \frac{f}{1+f} \tilde{\theta} + \frac{k^2}{4(1+f)} \delta_\DR + k^2 \psi
\,,\label{eq:tight}
\eeq
where $f \equiv 3 \bar{\rho}_2 / 4 \bar{\rho}_\DR$. As it must, this 
reduces to the equation for $\dot{\theta}_2$ or $\dot{\theta}_\DR$ 
\emph{without} the scattering term in the $f \to \infty$ (no DR) or $f 
\to 0$ (no $\chi_2$) limits, respectively. Since $f$ grows linearly in 
$a$, it will eventually be $\gg 1$ and the linearized evolution 
equations for $\chi_2$ will become identical to those for $\chi_1$. It 
follows that, in order to significantly modify the evolution of 
$\chi_2$, it is necessary to have $f \lsim 1$ when the mode enters the horizon. Recalling the parametrization~\Eq{r_def} and that $\Delta 
N_\text{eff} = \bar{\rho}_\DR / \bar{\rho}_\nu$, the value of $f$ at 
matter-radiation equality is given by
\beqa
f_\text{eq} 
= \frac{3 \bar{\rho}_2}{4 \bar{\rho}_\DR} \biggr|_\text{eq}
&= \frac34 \frac{r}{\Neff} \frac{\bar{\rho}_\DM}{\bar{\rho}_\nu} \biggr|_\text{eq}
\\
&\simeq \frac34 \frac{r}{\Neff} \frac{3}{1+\Omega_\B / \Omega_\DM} \Bigl (1 + \frac{\bar{\rho}_\gamma}{3\bar{\rho}_\nu} \Bigr|_\text{eq} \Bigr)
\,,
\eeqa
where we have approximated $\bar{\rho}_\DM \bigr|_\text{eq}$ as $\bar{\rho}_\gamma + 3\bar{\rho}_\nu \bigr|_\text{eq}$ since we know that the amount of extra radiation, $\Delta N_\text{eff}$, is constrained to be small.
Using $\Omega_\B / \Omega_\DM = 0.19$, the above expression gives
\beq
f_\text{eq} \simeq 0.2 \cdot \frac{r}{0.02} \cdot \frac{0.5}{\Neff} 
\,,
\eeq
where $r=0.02$ and $\Neff = 0.5$ are close to the benchmark values we 
will use later in our numerical study. To see whether $f_\text{eq}$ of 
$0.2$ is small enough to qualify as ``$f \ll 1$'', note that it is 
comparable to the value of $f$ for the standard baryon acoustic 
oscillation (BAO) case,
\beqa
f_\text{eq}^\text{\tiny BAO} 
= \frac{3 \bar{\rho}_\B}{4 \bar{\rho}_\gamma} \biggr|_\text{eq}
= \frac34 \frac{1}{1+\Omega_\DM / \Omega_\B} \Bigl (1 + \frac{3\bar{\rho}_\nu}{\bar{\rho}_\gamma} \Bigr|_\text{eq} \Bigr)
\simeq 0.2
\,.
\eeqa
It follows that the initial behavior of the $\chi_2$-DR system after 
matter-radiation equality is just like that of baryon-photon fluid, 
namely the density perturbations in $\chi_2$ undergo acoustic 
oscillations. In the meantime, the density perturbations in $\chi_1$ 
grow monotonically. Eventually, around recombination or shortly after, 
$f$ becomes $\gtrsim 1$ and the perturbations in $\chi_2$ begin to grow 
monotonically as well. This is where the analogy with the baryon-photon 
fluid breaks down, because the $\chi_2$-DR system never undergoes 
recombination. Once $\rho_\DR$ becomes subdominant to $\rho_2$, DR 
particles do not have enough energy to maintain dark acoustic 
oscillations. After this the equations for $\chi_2$ take the same form 
as those for collisionless CDM, and therefore its inhomogeneities in the linear regime evolve just like those of $\chi_1$. However, because 
$\chi_2$'s growth was initially held back for a while, we continue to 
have $\delta_2 \ll \delta_1$. At very late times, the behavior of 
$\chi_1$ and $\chi_2$ are expected to differ on small scales, since 
$\chi_1$ virializes while $\chi_2$ remains thermal.

Needless to say, the above conclusion applies only to the modes that 
enter the horizon before matter-radiation equality, and to a lesser 
extent also to the modes that come in between the equality and 
recombination. The modes that come in much later enter the horizon when 
$f \gg 1$, so they evolve just like $\chi_1$ from the beginning, leading 
to $\delta_2 = \delta_1$. This is a robust prediction of our scenario: 
\emph{the modes with $k \ll k_\mathrm{eq} \sim 0.01\Mpc^{-1}$ should see 
no modifications in the matter power spectrum compared to standard 
collisionless CDM\@.}

We are now ready to analyze how the suppression of $\delta_2$ affects 
the evolution of $\chi_1$ for the modes with $k \gg k_\mathrm{eq}$. At 
matter-radiation equality, these modes are already well within the 
horizon. Then, the relevant component of the perturbed Einstein 
equation,
\beq
k^2\phi + 3\frac{\dot{a}}{a} \!\left( \dot{\phi} + \frac{\dot{a}}{a} \psi \right)\!
= -4\pi G a^2 \delta \rho
\,,\label{eq:evolution:psi}
\eeq
just reduces to the Newtonian, Poisson equation, $k^2 \phi \simeq -4\pi 
G a^2 \delta\rho$. (The $a^2$ on the right-hand side is there because 
the $k$ on the left-hand side is the \emph{comoving} wave number 
related to the physical wave number $k_\text{phys}$ as $k_\text{phys} = 
k / a$, with $a=1$ today by convention.) Here, recalling the sum rule 
$\delta\rho = \sum_s \bar{\rho}_s \delta_s$, we have
\beqa
\delta \rho 
&= \bar{\rho}_1 \delta_1 + \bar{\rho}_2 \delta_2
\\
&= \bar{\rho}_\DM \bigl[ (1-r) \delta_1 + r \delta_2 \bigr]
\,,
\eeqa
where for simplicity we have only kept $\chi_1$ and $\chi_2$, while 
ignoring the other components such as baryons. (However, these are 
included in our numerical studies to be discussed later.) If $\chi_2$ 
were also collisionless CDM like $\chi_1$, we would have $\delta_1 = 
\delta_2 \equiv \delta_\DM$. Then the above relation would just lead to 
$\delta\rho = \bar{\rho}_\DM \delta_\DM$, as if we only had a single 
component of CDM\@. However, as discussed above, the $\chi_2$-DR 
interactions lead to $\delta_1 \gg \delta_2$ for the modes with $k \gg 
k_\mathrm{eq}$, so we instead have
\beq 
\delta\rho \simeq (1-r) \bar{\rho}_\DM \delta_1
\,.\label{eq:suppressed_delta_rho}
\eeq
This result is completely intuitive: if $\chi_2$ is not fluctuating, 
the fluctuations in the spacetime metric should only depend on the fluctuations in $\chi_1$,
but $\chi_1$ is only $(1-r)$ of the total mass density.
Then, the Poisson equation becomes
\beqa
k^2 \phi 
&\simeq -4\pi a^2 G \bar{\rho}_\DM \cdot (1-r) \delta_1
\\
&= -\frac{6}{\tau^2} \cdot (1-r) \delta_1
\,,
\eeqa
where we have used the Friedman equation and the fact that $a \propto \tau^2$ during matter domination.
Defining a dimensionless variable $\eta \equiv k\tau$, the above equation becomes
\beq
\phi \simeq -\frac{6(1-r)}{\eta^2} \delta_1
\,.\label{eq:phi_intermsof_delta1}
\eeq
On the other hand, cancelling $\theta_1$ between the two equations in~\Eq{evolution:DM1}, 
we get
\beq
\delta_1'' + \frac{2}{\eta} \delta_1' = -\psi + \frac{6}{\eta} \phi' + 3 \phi''
\,.
\eeq
where the primes denote derivatives with respect to $\eta$. Since the 
modes in question are well within the horizon, we have $\eta \gg 1$ by 
definition and the $\phi'$ and $\phi''$ terms can be neglected compared 
to the $\psi$ term. For simplicity, let us ignore the shear stresses in 
the photons and neutrinos, since our primary focus is on the difference 
between the $r=0$ and $r \neq 0$ cases. Then we have $\psi = \phi$, and so 
combining the above equation with Eq.~\Eq{phi_intermsof_delta1} leads to 
\beq \delta_1'' + \frac{2}{\eta} \delta_1' \simeq -\phi \simeq 
\frac{6(1-r)}{\eta^2} \delta_1 \,.
\eeq 
This tells us that $\delta_1$ is given by a linear combination of 
$\eta^{-1/2 \pm \sqrt{25/4\, - 6r}}$. The solution with the negative 
power of $\eta$ is important for matching the ``initial'' condition at $a=a_\text{eq}$, but 
then it quickly decays away. Then the late-time behavior of the modes of 
$\chi_1$ with $k \gg k_\mathrm{eq}$ is approximately given by 
\beq \delta_1 \propto 
\left(\frac{\eta}{\eta_\text{eq}}\right)^{\!\! -1/2 + \sqrt{25/4\, - 6r}} \propto 
\left(\frac{a}{a_\text{eq}}\right)^{\!\! 1 - 0.6r + \mathcal{O}(r^2)} \,. 
\eeq 
Taking the ratio of our case ($r > 0$) to 
standard, single-component, collisionless CDM ($r=0$), we obtain
\beq\label{eq:deltaratio} 
\frac{\delta_1 (r)}{\delta_1(0)} = \left(\frac{a}{a_\text{eq}}\right)^{\!\! -0.6r + \mathcal{O}(r^2)}\,, 
\eeq 
where we have neglected the mild $r$ dependence in the proportionality factor. This is justified 
because inflation gives us the same, universal initial conditions for 
the perturbations, irrespective of $r$ 
(see~Eqs.~\Eq{evolution:initial:RD} and~\Eq{evolution:initial:MD}), and 
the subsequent evolution of $\chi_1$ during radiation domination is 
controlled by radiation and hence is only mildly affected by the presence of 
$\chi_2$. We see from~Eq.~(\ref{eq:deltaratio}) that $\delta_1$ grows slower than in the CDM case.\footnote{Although 
the underlying dynamics is different, it is possible to get a similar 
effect by increasing the neutrino masses~\cite{Lesgourgues:2014zoa}.}

It follows that, in this approximation, the ratio of our power spectrum 
to the standard one is given by
\beq
\label{eq:powerlaw} 
\frac{P(r)}{P(0)} = \frac{[\delta\rho(r)]^2}{[\delta\rho(0)]^2} = 
\frac{(1-r)^2[\delta_1(r)]^2}{[\delta_1(0)]^2} \simeq (1-2r)
\! \left(\frac{a}{a_\text{eq}}\right)^{\!\! -1.2r} \,, 
\eeq 
where we have used the relation~\Eq{suppressed_delta_rho} in the second 
step and dropped $\mathcal{O}(r^2)$ terms at the end. As discussed 
already, the above suppression applies only to the modes with $k \gg 
k_\mathrm{eq}$, and the modes with $k \ll k_\mathrm{eq}$ see no 
suppression. The upper and lower panels of Fig.~\ref{fig:psratio} in \Sec{numerical} show the results of 
a numerical analysis that confirms this analytical understanding.

%

\section{Numerical results and corrections to the CMB spectrum}
\label{sec:numerical}
In this section, we obtain numerical results for the matter power 
spectrum in PAcDM framework. We also show that the effects of 
$\chi_2$-DR scattering on the CMB spectrum are small and within current 
experimental uncertainties.

The evolution of the perturbations in $\chi_{1,2}$ and DR is governed by 
Eqs.~\Eq{evolution:DM1}--\Eq{evolution:DR} and~\Eq{evolution:psi}, where 
the dark U(1) coupling is chosen to satisfy the condition~\Eq{Gamma>>H0}. 
The equations for baryons and photons are obtained simply by taking 
Eqs.~\Eq{evolution:DM2} and~\Eq{evolution:DR} and relabelling 
$\chi_2\to \text{B}$ (baryon) and $\text{DR}\to\gamma$, again in the tight 
coupling limit. The tight coupling approximation means our analysis 
fails to properly capture the physics of recombination and photon 
diffusion. We also ignore the effect of neutrinos. These limitations are 
not an immediate concern for us because we are only interested in how 
$\chi_2$-DR scattering affects the CMB spectrum \emph{compared to} the 
cases without such scattering, with or without DR\@, and the effects we 
ignore are common to all cases. Finally, photon polarizations are not 
distinguished in our treatment, but this is a small effect as we focus 
only on the matter and CMB temperature power spectra.

The tight coupling approximation and the absence of shear stress from 
neutrino free streaming imply that $\phi = \psi$. Assuming inflation, 
the initial conditions for the perturbations are then simply given by 
the superhorizon solutions of all the evolution equations above, together 
with the assumption of adiabatic perturbations that entropy per matter 
particle is unchanged by perturbations. For the modes that enter the 
horizon during radiation domination, the initial conditions are given by
\beq
\delta_{\gamma, \DR} = \frac43 \delta_{1,2,\text{B}} = -2\psi
\,,\quad
\theta_{1,2,\text{DR}, \text{B}, \gamma} = \frac{k^2 \tau}{2} \psi 
\,,\label{eq:evolution:initial:RD}
\eeq
while for those that come in during matter domination we have
\beq
\frac34 \delta_{\gamma, \DR} = \delta_{1,2,\text{B}} = -2\psi
\,,\quad
\theta_{1,2,\text{DR}, \text{B}, \gamma} = \frac{k^2 \tau}{3} \psi 
\,.\label{eq:evolution:initial:MD}
\eeq
We neglect the tilt in the primordial spectrum (i.e, $n_s = 1$) and use 
a $k$-independent value $10^{-4}$ for the initial perturbation $-2\psi$ 
above, but the precise choice of this number is immaterial as we will 
only compare the ratio of the spectrum with $r \neq 0$ to that with 
$r=0$.

We are now ready to evolve the perturbations numerically as governed by 
the equations and initial conditions described above, We choose the 
values $h=0.68$, $\Omega_{\gamma} h^2 = 2.47 \times 10^{-5}$, 
$\Omega_{\Lambda} h^2 = 0.69$, $\Omega_{b} h^2 = 2.2\times 10^{-2}$ and 
$\Omega_{\nu}=0.69 \Omega_{\gamma}$ \cite{Ade:2015xua}. $\Omega_\Lambda$ 
only has small effects on our results so its precise value is not important for 
our purpose here. We choose ${\Neff^\text{scatt}} = 0.4$ and a slightly 
larger value of $\Omega_\DM h^2 = 0.13$ in order to keep the redshift at 
matter-radiation equality unchanged. This allows us to compare our 
matter power spectrum to that of a conventional single component DM 
model \emph{without any} DR. We find the choice of $r=2.0\%$ leads to a $10\%$ 
suppression in the matter power spectrum around the scale $k \sim 0.2 
h\Mpc^{-1}$ compared to $\Lambda$CDM, thereby solving the $\sigma_8$ 
problem. It should be noted that this corresponds to a 
suppression of about $20\%$ 
compared to the $r=0$ case with \emph{the same amount} of DR, as 
shown in \Fig{psratio}, and confirms the features we identified 
analytically in \Sec{evolution}.\footnote{The detailed CMB and Large Scale Structure constraints on Partially-Interacting DM were studied in~\cite{Cyr-Racine:2013fsa} using the Planck 2013 data~\cite{Ade:2013zuv}. The bound allows $\lsim 5\%$ of the DM density 
to be tightly coupled to DR. Since we require an $r$ of order $2\%$, our result is consistent with this limit.}

The lower panel of Fig.~\ref{fig:psratio} shows the results with the same ratios $r$ of DM components, but with a reduced amount of DR, $\Neff^\text{scatt} = 0.05$. Notice that the suppression in $\sigma_8$ is nevertheless almost identical to the case with $\Neff^\text{scatt} = 0.4$. The essential reason why the reduction in $\sigma_8$ is so robust is that, even with $\Neff^\text{scatt}$ as small as $0.05$, there remains enough DR for $\chi_2$ to scatter with at the time when the $\sigma_8$ modes enter the horizon, which is well before matter-radiation equality. Therefore, our mechanism would still constitute a solution to the $\sigma_8$ problem even if future measurements were to settle on a smaller $\Neff^\text{scatt}$.

\begin{figure}
\begin{center}
\includegraphics[width=11.2cm]{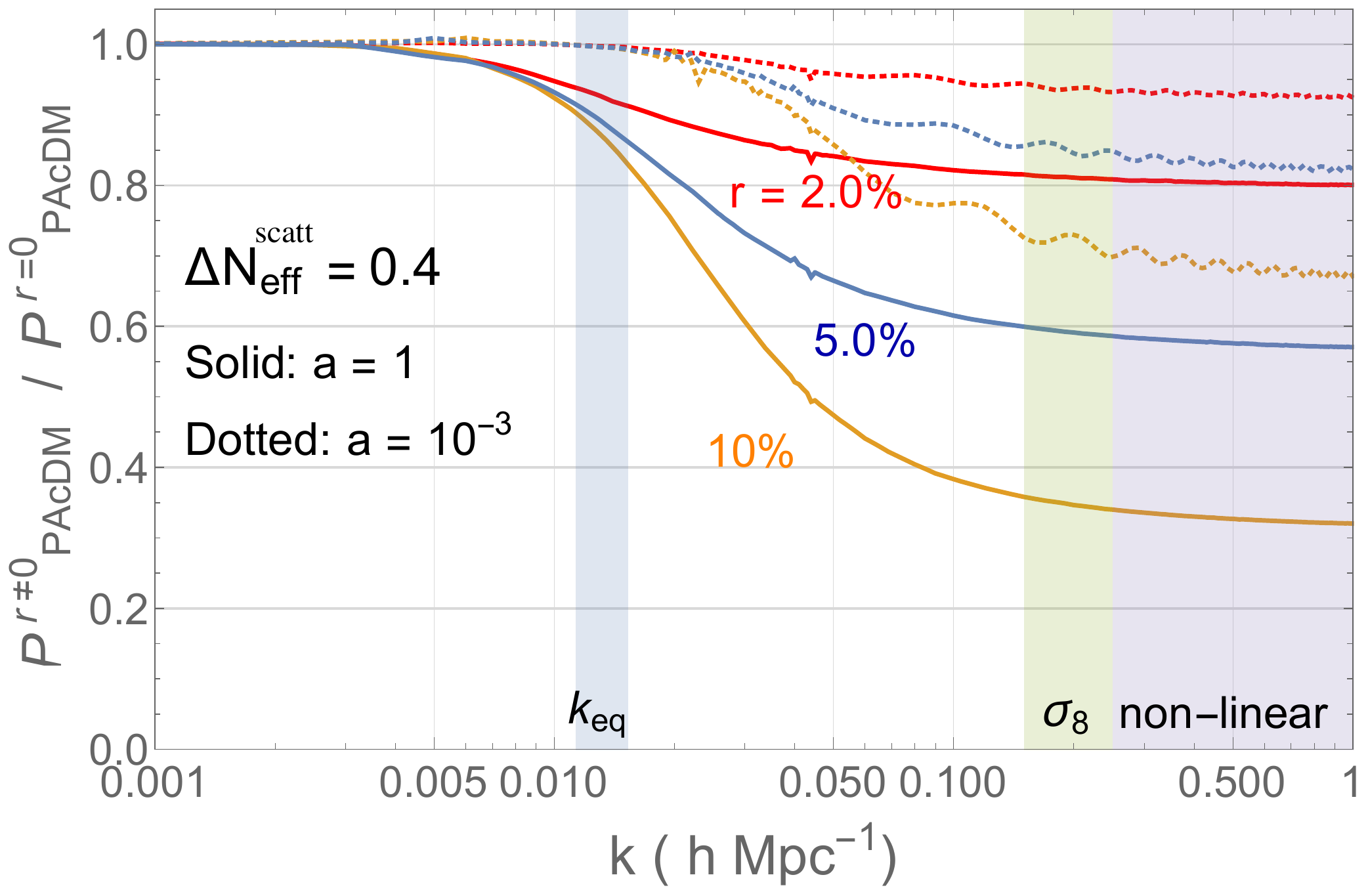}
\vspace{1.5em}
\\
\includegraphics[width=11.2cm]{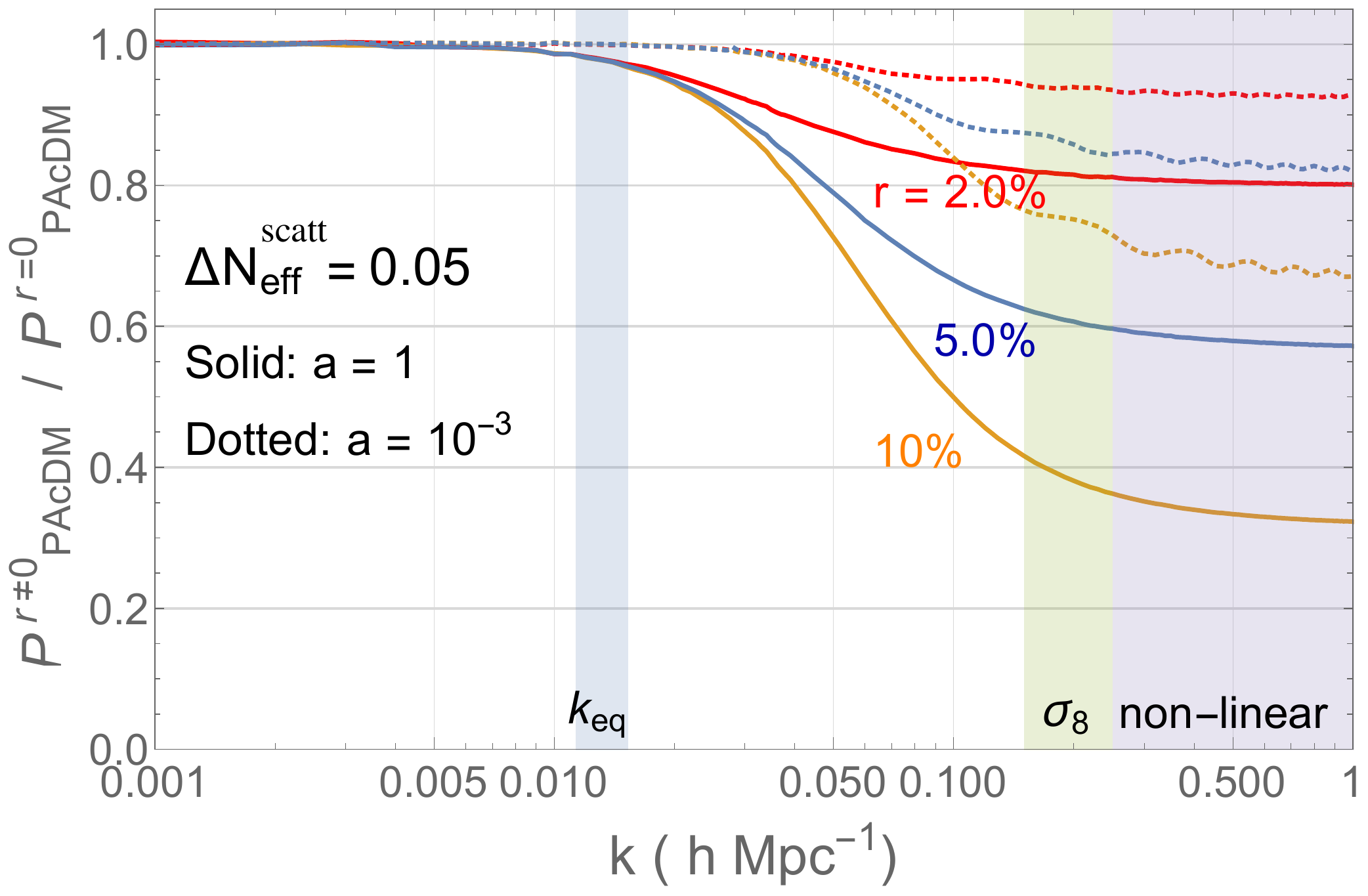}
\caption{Upper: Ratio of the DM power spectrum of the $r\neq 0$ case to the $r=0$ case, both with $\Neff^\text{scatt} = 0.4$.
The curves are obtained by numerically solving the linear evolution equations~\Eq{evolution:DM1}--\Eq{evolution:DR} and
the perturbed Einstein equation~\Eq{evolution:psi}, all in the tight coupling limit and assuming no anisotropic stress (hence $\sigma =
0$ and $\phi = \psi$).
Results for different values of $r$ are labelled in different colors,
while earlier ($a = 10^{-3}$) and later ($a=1$) times are indicated by dotted and solid lines, respectively.
For the smaller scale structures $k \gtrsim 0.2h\Mpc^{-1}$, nonlinear gravitational effects become important so our linear approximation
is no longer reliable. Lower: Same plot but with a reduced amount of DR, $\Neff^\text{scatt} = 0.05$.
}\label{fig:psratio}
\end{center}
\end{figure}

Let us now discuss the impact of $\chi_2$-DR interactions on the CMB 
spectrum. We compare the temperature power spectra of the $r=2.0\%$ and 
$r=0$ cases with the same amount of DR\@. Since our equations do not 
include the physics of recombination or photon diffusion, we 
halt the evolution just before recombination at $a=10^{-3}$ (which is 
when the electron number density begins to fall exponentially) 
and then evaluate the following quantity:
\beq
\!\left( \frac{\delta T}{T} \right)_{\!\!*}
\equiv \frac14 \delta_\gamma + \psi
\,.
\eeq
This has the same form as the standard expression for $\delta T / T$ for the 
CMB, except for the absence of small Doppler effect corrections~\cite{Hu:2008hd} and 
the fact that it is evaluated right before recombination as opposed to the time of photon decoupling. 
Nevertheless, we map $(\delta T / T)_*$ to the $C_\ell$ 
coefficients as if $(\delta T / T)_*$ were $\delta T / T$. In 
other words, we treat the system as if the photons instantaneously 
decouple right before recombination, and obtain a ``snapshot'' of the 
CMB spectrum at that time. This is sufficient for the purpose of showing 
that the effects of $\chi_2$-DR interactions have very small impact on 
the CMB spectrum, because the effects of such interactions on the photons at and after recombination are small. Moreover, there is very little time 
between recombination and photon decoupling. Hence, if the CMB spectrum 
with $r = 2.0\%$ is very similar to that with $r=0$ right before 
recombination, we expect that they should continue to remain similar 
afterwards.
In \Fig{CMB}, we compare the primordial spectra of the $r \neq 0$ (red 
solid) and $r=0$ (black solid) cases. The $\chi_2$-DR scattering damps 
the CMB spectrum, but the deviations are always less than $2\%$, which 
is within the uncertainties of current measurements. The largest deviations are for modes in the range $700\lsim\ell\lsim1600$. The reason that the 
modes with $\ell\lsim 700$ see very little deviation is because they 
enter the horizon late and by that time the DR has become too cold to affect 
the physics, as discussed in \Sec{evolution}. On the other hand, the 
modes with $\ell \gsim 1600$ enter the horizon at a time when the 
energy density is completely dominated by radiation and so the metric 
perturbations damp quickly, resulting in small corrections to $\delta_{\gamma}$. 

\begin{figure}[t!]
\begin{center}
\includegraphics[width=12cm]{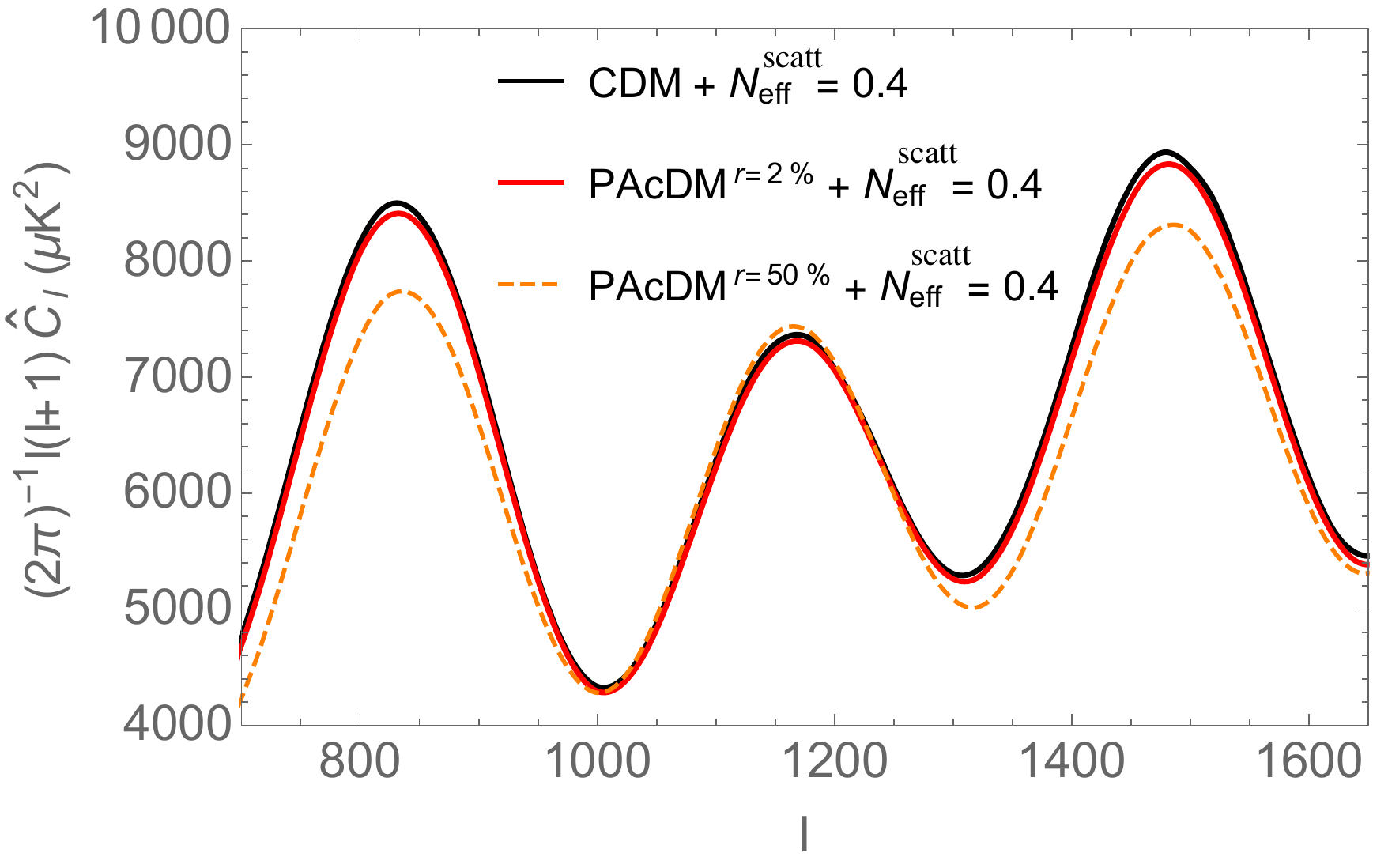}
\caption{A comparison of the CMB spectrum between PAcDM and CDM models, assuming 
$\Neff^\text{scatt} =0.4$ in both cases. The black (red) curve is for the $\Lambda$CDM (PAcDM) model, 
derived from the $\{\delta_{\gamma},\theta_{\gamma}\}$ result in the linear  evolution equations. 
The PAcDM model assumes the DM ratio $r=2.0\%$. For comparison, we also show a PAcDM model with $r=50\%$, which exhibits a clear enhancement of the expansion peaks and suppression of the compression peaks due to the pressure of tightly coupled DR.}
\label{fig:CMB}
\end{center}
\end{figure}

 The other aspect of the last point above is that the metric 
perturbations are appreciably modified only after matter-radiation 
equality. This means that the observations of CMB lensing effects can porentially 
constrain our scenario. The key quantity for calculating the impact of 
lensing on the temperature power spectrum is the lensing power spectrum 
$C_\ell^{\phi\phi}$ as a function of multipole $\ell$. Using the Limber 
approximation \cite{Limber}, we estimate the spectrum 
using~\cite{Pan:2014xua}
\beq
\ell^4 C_\ell^{\phi\phi}
\propto
\int_0^{\chi_*}\!\! \dd\chi \, \bigl( 1-\chi / \chi_* \bigr)^{\! 2} \, k \phi^2(k,a) \, g^2(\chi)
\,,\label{eq:CL}
\eeq
 where $\chi$ is the comoving distance from the observer and hence $k = 
\ell / \chi$. In this expression the scale factor $a$ is to be evaluated at $\chi$, i.e., $a 
= a(\chi)$. The parameter $\chi_* \simeq 10^{4}\Mpc$ corresponds to the 
value of $\chi$ of 
the last scattering surface. The lensing effect is captured by $(1-\chi 
/ \chi_*)^2$. The function $g(\chi)$ effectively describes the decays of 
metric perturbations due to the vacuum energy at late times. Its value 
is therefore equal to $1$ by definition during matter domination ($\chi \gsim 0.5 
\chi_*$) and starts to decrease once $\chi$ drops below $\simeq 0.5 
\chi_*$. The transfer function is already folded into the above 
expression, so $\phi(k)$ is simply the primordial metric fluctuation 
determined by the linear perturbation equations discussed in 
\Sec{evolution}.

In the Planck data~\cite{Ade:2015zua}, the smallest percentage error in 
$C_\ell^{\phi\phi}$ is $\simeq 5\%$ at $\ell \simeq 150$. The 
$C_\ell^{\phi\phi}$ at this multipole gets its main contributions from 
$\chi \simeq 0.4 \chi_*$ in Eq.~\Eq{CL}, which corresponds to the modes 
with $k \simeq \cO(0.01)\Mpc^{-1}$. The deviation in $\phi^2$ between 
the $r =2.0\%$ and $r=0$ cases is $\simeq 4\%$ for this value of $k$ 
(see \Fig{psratio}), and the integral~\Eq{CL} gives a $2.5\%$ deviation 
between the two cases. This is within the current uncertainties.

\section{Conclusions}
\label{sec:conc}
 We have presented a new framework in which DM is composed of two 
distinct components that can provide a solution to both the $H_0$ and 
$\sigma_8$ problems. While the dominant component of DM is cold and 
collisionless, the subdominant component is also cold but interacts 
strongly with DR, which itself constitutes a tightly coupled fluid. Our 
framework is very general and can be adopted in a wide variety of DM 
models. In particular, it can easily be accommodated in the hidden WIMP 
framework, with both constituents of DM arising as thermal relics. 
Our scenario predicts distinctive modifications to the matter and CMB 
power spectra, allowing it to be tested by future 
experiments.

By solving a set of linear evolution equations, we have shown that the 
observed $10\%$ discrepancy in $\sigma_8$ requires the mass density in 
the subdominant, interacting DM species to constitute $\simeq 2.0\%$ of 
the total DM density, while the amount of DR can be separately chosen to 
fix the $H_0$ problem. This apparently small ratio of the two DM 
components could easily arise in, for example, the WIMP framework 
without introducing hierarchically small parameters into the Lagrangian. 
The required tight couplings between the interacting DM and DR and 
within DR itself can be obtained in a wide range of perturbative 
coupling constants, as we showed in a concrete model. We found that, 
with an interacting DM component of about $2\%$ to solve the $\sigma_8$ 
problem and the appropriate amount of DR to address the $H_0$ problem, 
the deviations in the CMB spectrum and the CMB lensing measurements are 
well within the current uncertainties. 

It is interesting to compare and contrast our proposal with the scenario put forward in~\cite{Buen-Abad:2015ova, Lesgourgues:2015wza, Ko:2016uft}.
In the PAcDM framework, only a subcomponent of DM experiences acoustic oscillations, while the primary component of DM is responsible for building up structure. As seen in Eq.~(\ref{eq:powerlaw}), this suppresses the rate of growth of power during the era of matter domination, with the result that most of the corrections to the DM density perturbations, 
and hence corrections to the gravitational potential, arise well after 
matter-radiation equality. However, in the proposal of~\cite{Buen-Abad:2015ova, Lesgourgues:2015wza, Ko:2016uft}, the entirety of DM undergoes oscillations prior to matter-radiation equality that continue through to the CMB epoch. In this case, the corrections to the DM density perturbations are already 
significant at the time of matter-radiation equality, and the resulting
corrections to the CMB are expected to be significantly larger. Hence future 
precision studies of the CMB may be able to distinguish these two 
classes of models.

It is important to note that our mechanism to reduce $\sigma_8$ is not especially sensitive to the precise value of $\Neff^{\text{scatt}}$. Hence, if future measurements were to settle on a smaller $\Neff^{\text{scatt}}$, our mechanism would still constitute a solution to the $\sigma_8$ problem. Note that lowering $\Neff^{\text{scatt}}$ would also imply even smaller corrections to the CMB, since most of the modes that are observed in the CMB would now enter the horizon at a time when the contribution of DR to the energy density is small.

While the primary focus of this article is on large scale structure, our 
framework may also have potentially observable effects on smaller 
scales. It follows from Eq.~\Eq{Gamma>>H0} that at each locale in the 
universe the $\chi_2$ particles continue to experience a sufficient 
number of collisions with surrounding DR particles to maintain thermal 
equilibrium, at least locally, at all times. This remains true during 
the era of structure formation. The same condition also ensures that the 
DR itself remains a tightly coupled relativistic fluid at all times. 
Being strictly massless, the dark charged particles in DR never undergo 
recombination. Being tightly coupled, DR is non-dissipative and hence 
behaves as a perfect thermal fluid. Because of these properties, which 
are qualitatively different from those of the baryon-photon system, we 
expect that the $\chi_2$-DR system does not collapse into a disk but 
instead forms a smooth, spherical halo around the galactic center. This 
is qualitatively distinct from the dynamics of the recently proposed 
``double-disk DM'' or partially dissipative DM~\cite{Fan:2013yva}. The 
existence of this halo would impact galactic dynamics, with the exact 
nature of its effects depending on the details of its density profile. 
We defer a careful study of these effects on the small scale structure of DM halos for future work.

The current discordance between direct and indirect measurements of 
$H_0$ and $\sigma_8$ may be the first hint for new cosmology beyond the 
$\Lambda$CDM paradigm. These discrepancies can be naturally addressed 
within the non-minimal dark sector structure we have proposed. In the 
coming years, the experimental precision in the indirect measurement of $N_{\rm 
eff}$ and $\sigma_8$ from the CMB~\cite{Abazajian:2013oma, 
Errard:2015cxa, Wu:2014hta, Dodelson:2016wal} (e.g. CMB stage-IV), and 
in direct measurements of the Hubble constant~\cite{Macri:2006wm, 
Greenhill:2009yi}, and $\sigma_8$~\cite{Hannestad:2007fb, 
Dodelson:2016wal} (LSST, DESI) are all expected to improve significantly. If 
the current discrepancies in the $H_0$ and $\sigma_8$ measurements are 
indeed due to new physics, these future experiments have great potential 
for distinguishing between different candidate theories such as the framework presented in this paper.

\begin{acknowledgments}

We thank Gustavo Marques-Tavares, Valentina Prilepina, Matthew Reece, Martin Schmaltz, 
Neelima Sehgal and Yong Tang for helpful discussions. ZC, SH and YT are 
supported in part by the National Science Foundation under grant 
PHY-1315155, and by the Maryland Center for Fundamental Physics. SH is 
also supported in part by a fellowship from The Kwanjeong Educational 
Foundation. YC is supported by Perimeter Institute for Theoretical 
Physics, which is supported by the Government of Canada through Industry 
Canada and by the Province of Ontario through the Ministry of Research 
and Innovation. YC is also supported in part by the Maryland Center for 
Fundamental Physics. TO is supported by the US Department of Energy 
under grant DE-SC0010102. YC, TO, and YT all thank the Aspen Center for 
Physics, which is supported by National Science Foundation grant PHY-1066293.

\end{acknowledgments}

\bibliographystyle{utphys}

\bibliography{./DMRef.bib}

\end{document}